\documentclass[a4paper]{article}
\usepackage{multirow}
\usepackage{INTERSPEECH2020}

\title{Speaker conditioned acoustic-to-articulatory inversion using x-vectors}
\name{Aravind Illa, Prasanta Kumar Ghosh}
  \address{Electrical Engineering, Indian Institute of Science (IISc), Bangalore-560012, India}
\email{aravindi@iisc.ac.in, prasantg@iisc.ac.in}

\begin{document}
\maketitle
\begin{abstract}
   Speech production involves the movement of various articulators, including tongue, jaw, and lips. Estimating the movement of the articulators from the acoustics of speech is known as acoustic-to-articulatory inversion (AAI). Recently, it has been shown that instead of training AAI in a speaker specific manner, pooling the acoustic-articulatory data from multiple speakers is beneficial. Further, additional conditioning with speaker specific information by one-hot encoding at the input of AAI along with acoustic features benefits the AAI performance in a closed-set speaker train and test condition.
   In this work, we carry out an experimental study on the benefit of using x-vectors for providing speaker specific  information to condition AAI. Experiments with 30 speakers have shown that the AAI performance benefits from the use of x-vectors in a closed set seen speaker condition. Further, x-vectors also generalizes well for unseen speaker evaluation.    
\end{abstract}

\section{Introduction}
Speech acoustics is a result of movements of the articulators namely tongue, lips, jaw, velum  which form constriction in the vocal tract \cite{goldstein2003articulatory}. Along with the speech acoustics, having the knowledge of articulatory position has been shown to benefit many speech applications including, multimedia \cite{ref3,ref4,ref5}, speaker verification \cite{speakerID}, 
automatic speech recognition \cite{ref9} and speech synthesis \cite{illa2019investigation,cao2017integrating}.
One of the state-of-the-art devices to acquire the synchronous acoustic-articulatory data is electromagnetic articulograph (EMA). In case where direct articulatory measurement is not feasible, to estimate articulatory movements, an acoustic-articulatory mapping is typically learned with the available acoustic-articulatory data.  The estimation of the articulatory movement from the speech acoustics is known as acoustic-to-articulatory inversion (AAI). In the literature various models were proposed for AAI, e.g., 
code-book based \cite{ref11},
Gaussian Mixture Model (GMM) \cite{ref12}, Hidden Markov Model (HMM) \cite{ref14} and neural network based approaches \cite{ref13, BLSTM, illa2018low}. The state-of-art performance is achieved by long short term memory (LSTM) networks, which is a recurrent neural network (RNN)  \cite{BLSTM, illa2018low}.



In order to learn the weights of an LSTM, one needs a significant amount of acoustic-articulatory data. However, collecting large amount of 
data using EMA from a speaker is impractical and cumbersome as sensors fall off in a long recording, and re-attaching them becomes a challenge causing discomfort to the subject.
To reduce the demand on the amount of acoustic-articulatory data from a speaker, a low resource AAI model has been proposed using 
generic model AAI (GM AAI) model \cite{illa2018low} which is trained by pooling the data from all speakers using bi-directional long short term memory networks (BLSTM).
The performance of the GM AAI is shown to be better than that of separate speaker dependent AAI (SD AAI) model, where training is performed   
using acoustic-articulatory data from each speaker separately. 
This indicates that BLSTM networks are able to capture multiple speakers' acoustic-to-articulatory mappings without drop in performance compared to the speaker dependent AAI models.
The GM AAI model can be fine-tuned with 
speaker specific acoustic-articulatory data, which results in further improvements in the performance of AAI \cite{illa2018low} compared to GM AAI.
In summary, this approach involves two-step training procedure, first to build GM AAI with multiple speakers' acoustic-articulatory data, and then fine-tuning GM AAI to speaker specific data. This results in separate AAI models for each speaker after speaker specific fine-tuning, which, although improve AAI performance, increases the number of models and, hence, the number of parameters, which, in turn, increases the storage requirement unlike the GM AAI model.

To overcome this limitation, an alternative approach to fine-tuning GM AAI is proposed, in \cite{SCOHv}, by conditioning BLSTM with auxiliary features which carry speaker specific information, along with the acoustic features for learning rich acoustic-to-articulatory mappings of multiple speakers within a single model.
This is performed by representing speaker specific information as one-hot encoding to condition GM AAI, known as speaker conditioned AAI (SC AAI) \cite{SCOHv}. It has been shown that SC AAI is efficient and provides a more  compact way of learning multiple AAI mappings compared to fine-tuning GM AAI for every speaker. But representing speaker specific information as one-hot encoding limits the SC AAI model to the closed-set speaker condition (train and test sets comprising the same set of speakers).
In this work, instead of representing speaker specific information with one-hot encoding, we utilize x-vectors which are known to encode the speaker specific information \cite{XV2018ICASSP}. We hypothesize that x-vector based SC AAI model (xSC AAI) would have following benefits:
i) it can achieve performance on par with the SC AAI in closed-set condition and can capture multiple speakers' AAI mapping within a single model,
ii)  unlike SC AAI, x-vectors can able to generalize in unseen speaker evaluation, where train and test speakers are different. 
In this work, we perform an experimental study to compare the performance of the proposed xSC AAI model with different baseline AAI methods in both seen and unseen speaker evaluation conditions.
Experiments are performed with acoustic-articulatory data of 30 speakers each speaking 460 English sentences, where 20 speakers are used for seen speaker AAI model evaluations and 10 speakers for unseen speaker evaluations.
Experimental results revealed that the xSC AAI model performs on par with one-hot representation based SC AAI in seen case evaluation, and outperforms the baseline methods in unseen speaker evaluation.


\section{Dataset}
For experiments, 460 English sentences from MOCHA TIMIT \cite{Mocha460} were chosen as a speech stimuli.
Acoustic-articulatory data was collected from 30 speakers, comprising 15 male (M1--M15) and 15 female (F1--F15) speakers.
All the speakers were proficient in English and reported to have no speech disorders in the past. 
Acoustic-articulatory data was collected using AG501 electro-magnetic articulograph (EMA) \cite{AG501}.

\begin{figure}[htb]
  \centering
    \vspace{-.4cm}
  \centerline{
  \includegraphics[trim = 7mm 214mm 35mm 7mm, clip, width=6cm]{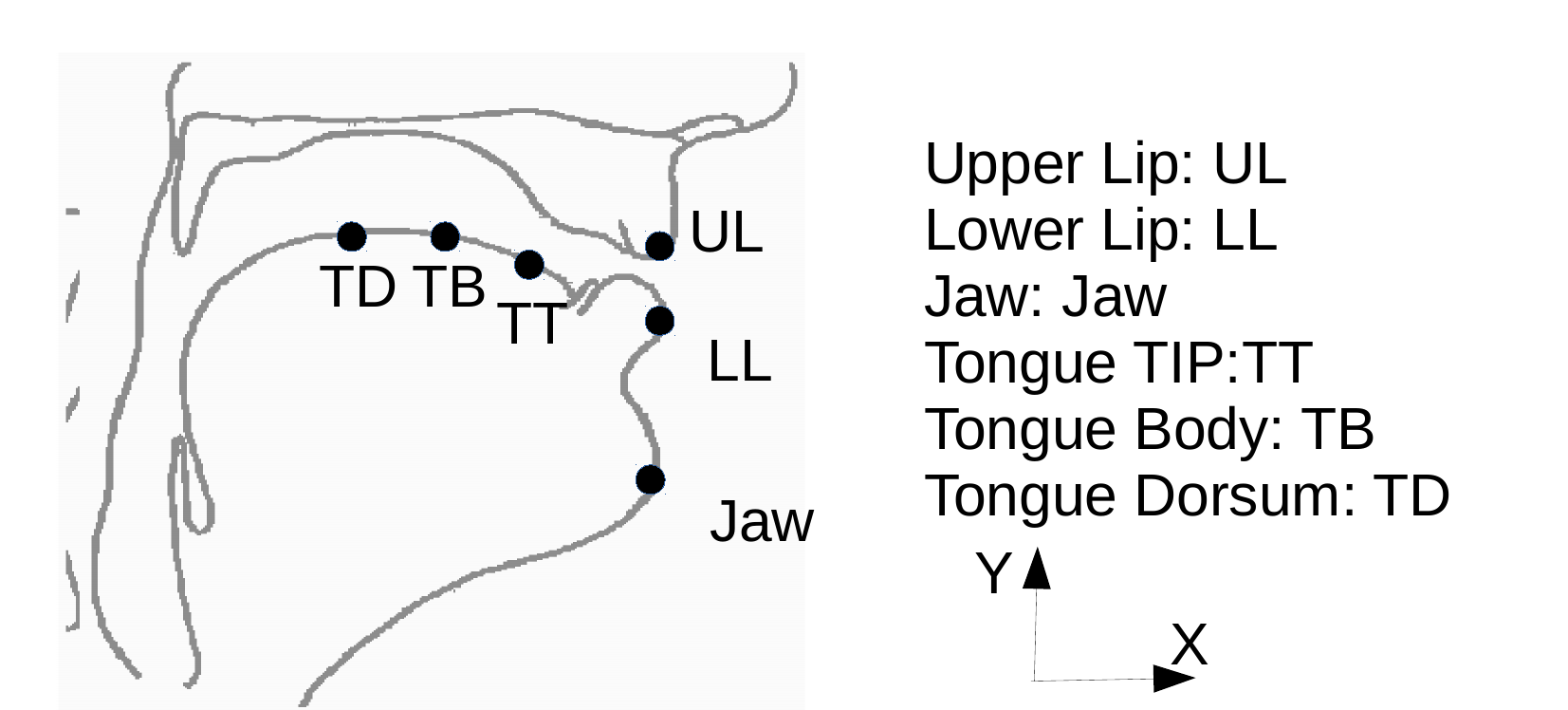}}
  	\caption{Schematic diagram indicating the placement of EMA sensors \cite{illa2018low}}
  	\label{EmaSetUp}
 \vspace{-.8cm}
 \end{figure}

Using the t.bone EM9600 shotgun microphone \cite{EM9600} placed in front of the speaker, acoustic data  was recorded at a sampling rate of 48kHz. Along with the acoustics, synchronous articulatory movement data was collected at 250Hz, by gluing (using ``Epiglu" \cite{GlueEMA})
sensors of AG501 on six
articulators namely upper lip (UL), lower lip (LL), jaw (Jaw) tongue tip (TT), tongue body (TB) and tongue dorsum (TD), following recommendations reported in \cite{optimal}. 
For head movement correction, two additional sensors were glued on the mastoids.
Fig. \ref{EmaSetUp} shows a midsagittal view of the vocal tract illustrating the placement of sensors, which capture the movements in horizontal and vertical directions indicated by X and Y. 
The sensors capture the articulatory movements in horizontal and vertical directions indicated by X and Y respectively in the  midsagittal plane \cite{illa2018low}. 
This results in a 12-dim articulatory feature vector which are indicated by  $UL_x$, $UL_y$, $LL_x$, $LL_y$, $Jaw_x$, $Jaw_y$, $TT_x$, $TT_y$, $TB_x$, $TB_y$, $TD_x$, $TD_y$.
During recording, each sentence was projected on a computer screen placed in front of the speaker, and a slide changer was
provided to the speaker to navigate through all the sentences. 
We recorded simultaneous acoustic-articulatory data for each sentence.
To remove the start and end silence segments in each sentence, we performed manual annotations to the recorded acoustic-articulatory data. This resulted in a total of 11.19 hours of acoustic-articulatory data with an average duration of  22.38 ($\pm$ 2.48) minutes per speaker.
 
\label{s2:dataset}
\section{Proposed approach}
\label{s3:scaai}

The mapping function from acoustic features to articulatory movements is known to be non-unique and nonlinear in nature \cite{ghosh2010generalized}.
Further, the relation between acoustic features and articulatory movements is not instantaneous due to co-articulation. Rather, the articulatory position at a time depends on the acoustics before and after that time instant. Also, it is known that the articulatory movement trajectories are smoothly varying in nature \cite{ghosh2010generalized}.
In order to learn this non-linear and complex function, neural networks have been shown to perform well in modeling the AAI. Recently, bi-directional long
short term memory (BLSTM) networks are shown to achieve the state-of-the-art performance for AAI task. BLSTM networks have also been shown to preserve the smoothness characteristics in the estimated articulatory trajectories \cite{BLSTM, illa2018low} implicitly and do not need any further post-processing steps for smoothing, unlike DNN and GMM. 
Recently using BLSTM, in \cite{SCOHv}, it has been shown that by conditioning BLSTM networks using speaker information (as one-hot encoded representation) would benefit the performance of AAI in a closed-set speaker training and evaluation condition.
To generalize the SC AAI to an unknown speaker, instead of condition the SC AAI model with the one-hot representation of speaker specific information, we, in this work, perform experiments with x-vectors as a representation to encode speaker information. 

x-vectors are neural network embeddings, which are recently shown to be successful in extracting speaker representations and widely used in speaker verification, language identification, and speaker diarization applications \cite{XV2020stateCSL}.
From the variable-length acoustic segments, the x-vector model computes speaker embedding using Time-Delay Deep Neural Network (TDNN) architecture \cite{XV2018ICASSP}. 
The first few layers in TDNN comprises time-delay network which are equivalent to dilated convolutions. The sequence of time-delay layers are used to extract information from the input frame-level features by aggregating context from previous and future frames.
The time-delay layers are followed by a pooling layer which computes the mean and standard deviation from the TDNN output over time which are used as a speaker embedding computed based on an input utterance from a speaker, which is known as x-vector. 


   \begin{figure}[htb]
   \centering
   \vspace{-.28cm}
   \centerline{\includegraphics[trim = 2mm 125mm 3mm 15mm, clip,width=\columnwidth]{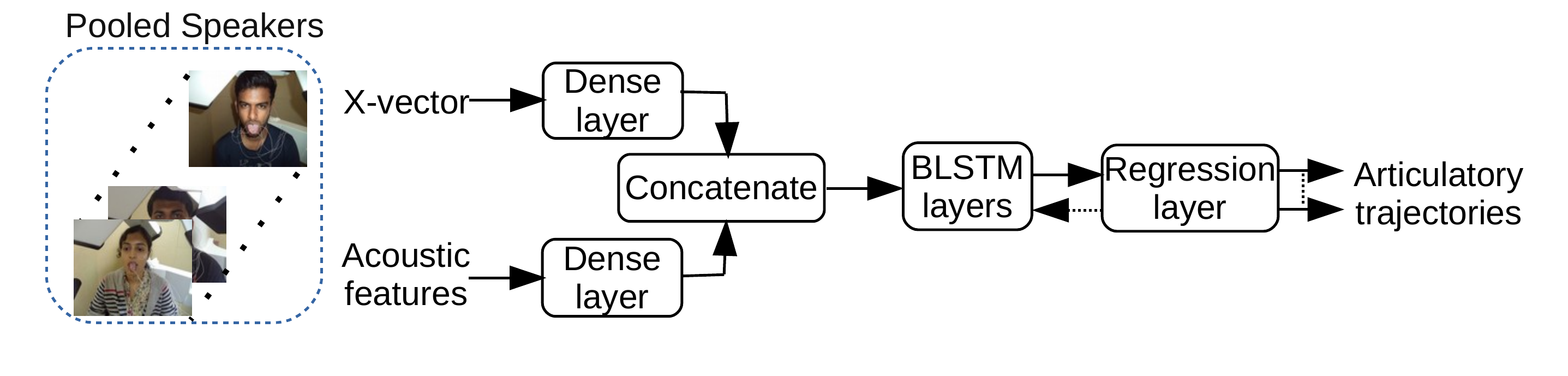}} 
    \vspace{-.35cm}
    \caption{Block diagram of the proposed SC AAI model using $\mathrm{x}$-vector approach ($\mathrm{x}$SC AAI).}
    \label{scaai}
    \vspace{-.4cm}
 \end{figure}

The proposed approach for SC AAI using x-vector, indicted as xSC AAI is illustrated in Fig. \ref{scaai}.
From a given speech utterance, we compute an acoustic feature vector at every speech frame (frame rate of acoustics features and the sampling rate of articulatory features are matched to obtain one to one correspondence between the two).
Further, from the variable length utterance, a fixed dimensional (512) x-vector is extracted using TDNN, and we replicate the  x-vector for every frame in the utterance to match the number of acoustic feature vectors. The acoustic features and x-vectors are fed to separate dense layers as shown in Fig. \ref{scaai}. The outputs of dense layers are concatenated and fed to the BLSTM layers.
The output of BLSTM layers is fed to a time-distributed linear regression layer using linear activation.  
The mean squared error between the original and predicted articulatory features is used as a loss function to learn the neural network weights.
Training of the xSC AAI model is performed with multiple speakers' acoustic-articulatory data.
In seen case evaluation, test sentences (non-overlapping with training sentences) are from the training speaker set. In this work, we choose 20 speakers for the seen case experimental set-up. While in unseen evaluation, neither acoustic features nor x-vectors from test speakers are utilized during the training of the SC AAI model. In this work, we have chosen 10 unseen speakers.
\section{Experimental setup}
\label{s4:exps}
\textit{Data post-processing and feature computation:}
We perform post-processing operations on the recorded acoustic-articulatory data, before computing acoustic and articulatory features.
The articulatory data is first low-pass filtered with a cutoff of 25Hz, to avoid high-frequency noise. This is done also because it is known that the articulatory trajectories are slowly varying in nature and most of the energy lies below 25Hz \cite{ghosh2010generalized,illa2020impact}.
Also, the articulatory data obtained at 250Hz is down-sampled to 100Hz make it synchronous with acoustic features.
As acoustic features, we compute 13-dim Mel-frequency cepstral coefficients (MFCCs) \cite{htk} from the recorded speech signal using a window size of 20ms with 10ms frameshift.
In the literature, MFCCs have been shown to be optimal for the AAI task using maximal mutual information criterion \cite{ghosh2010generalized} as well as using representations learned from the raw waveform \cite{illa2019representation}.
At an utterance level, we perform mean and variance normalization of both the acoustic and articulatory features.
From the 30 speakers, as described above, we choose 20 speakers  (10 male: M1-- M10 and 10 female: F1-- F10) for the train and test sets under seen speaker condition and 10 speakers (5 male: M11 -- M15  and 5 female: F11 -- F15) for unseen speaker evaluation.
For all the experiments, from each speaker (both seen and unseen) the recorded acoustic-articulatory data using 460 sentences were divided into a training set 80$\%$ (364), validation set 10$\%$ (46) and testing set 10$\%$ (46).



\textit{AAI schemes and evaluation metrics:}
In order to compare the performance of the xSC AAI, we consider baselines with different AAI models.
In seen speaker evaluation, we consider 
different training schemes resulting in the following AAI models: a) speaker dependent AAI model (SD AAI): train and test in a speaker specific manner using same speaker acoustic-articulatory data,
 b) Generic Model AAI model (GM AAI): one single AAI model is trained and tested by pooling data from all 20 speakers,
c) GM AAI with speaker specific fine-tuning (GM-FSD AAI): the GM AAI model is further fine-tuned with speaker specific acoustic-articulatory data, which results in an AAI model for each speaker,
d) Speaker condition using one-hot vector (SC AAI): one single AAI model is trained similar to the GM, but we provide auxiliary speaker specific information using a 20-dim one hot vector along with the acoustic features.
e) Speaker condition using x-vector (xSC AAI): In the proposed approach we condition AAI with x-vectors while training with 20 speakers' acoustic-articulatory data.
 
 In unseen speaker evaluation, we use the models trained with 20 speakers acoustic-articulatory data but for testing, we use sentences from 10 unseen speakers.
 We consider the following models: a) GM and xSC AAI models: trained with 20 speakers (F1-F10; M1-M10) tested with 10 speakers (F11-F15; M11-M15), b) uSC AAI: as a baseline, we use the SC AAI model trained with 20 speakers data.  However for testing SC-AAI with unseen speakers, one-hot representation is not possible. So, we trained a speaker identification network (SID) with 20 speakers' acoustic data. For a test sentence of an unseen speaker, we take softmax output of the SID network as a speaker identity vector for SC AAI models. We hypothesize that by this approach (uSC AAI), we could represent an unseen speaker as a combination of speakers used for training SC AAI and SID networks.
\begin{figure}[htb]
  	\centering
  	  \vspace{-.28cm}
  	\centerline{\includegraphics[trim = 2mm 5mm 10mm 10mm, clip,width=7cm]{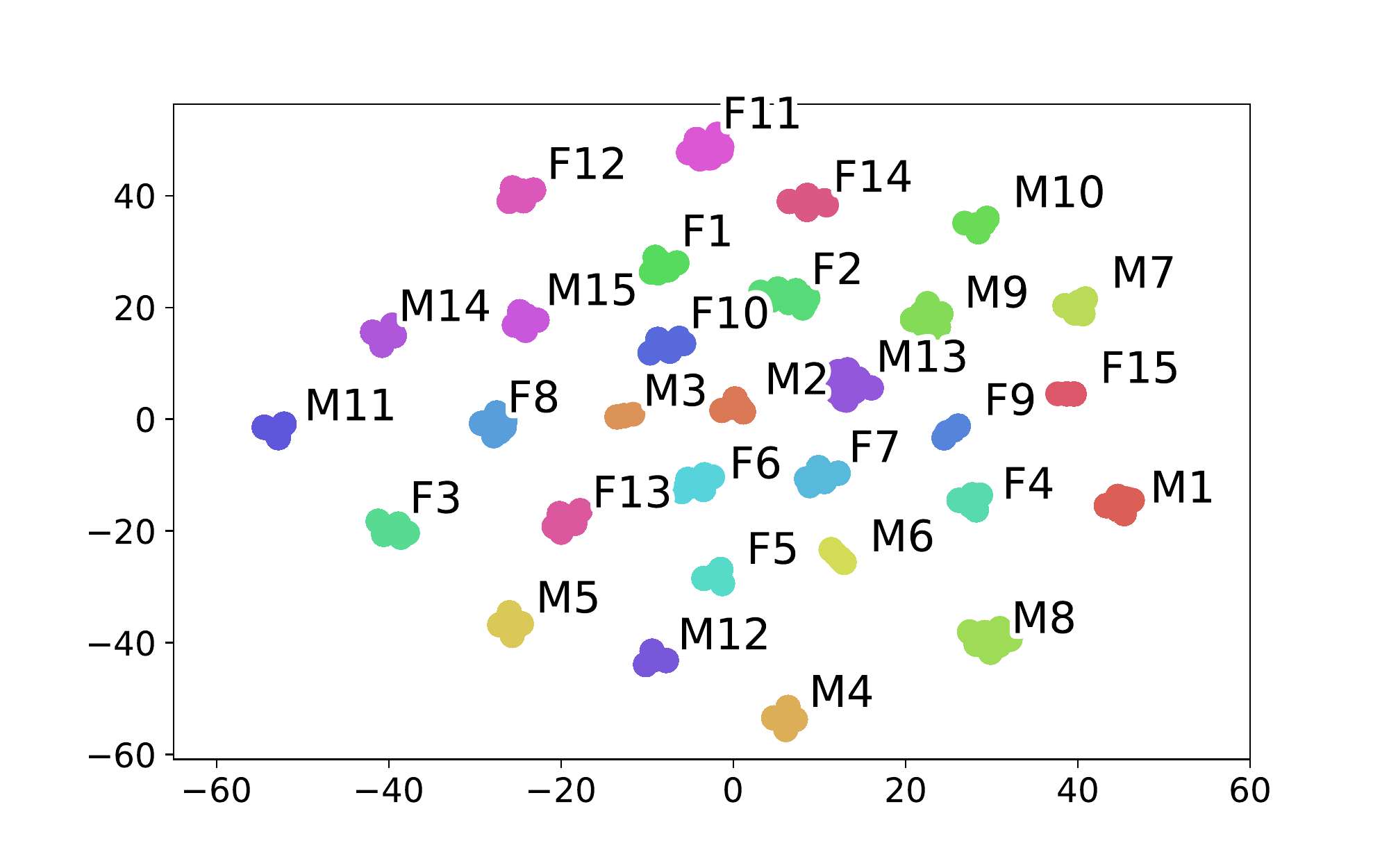}} 
  	\vspace{-.35cm}
  	\caption{Visualizations of $\mathrm{x}$-vector speaker embeddings using t-SNE for 30 speakers used in this work.}
  	\label{tsne}
  	\vspace{-.4cm}
  \end{figure}

\textit{Hyper-parameter details and neural network training:}
To perform experiments with xSC AAI, we choose a dense layer with 200 units which takes 13-dim MFCC as an input.
For speaker embedding layer, where input is a 512-dimensional x-vector, we choose a dense layer with 32 units. For the BLSTM, we use 3 hidden layers with 256 dim output units in each layer.
The BLSTM layer output is fed to a time-distributed linear regression layer with 12-dim output.  For training, we minimize mean squared error, and early stopping is performed based on the validation loss.
x-vectors are computed using Kaldi toolkit \cite{povey2011kaldi}. We used a pre-trained model trained on the VoxCeleb database following the recipe ``egs/voxceleb/v2" in Kaldi \cite{VoXcelb}. 
To assess the discriminative ability of the extracted x-vectors across all 30 speakers,  Fig. \ref{tsne} illustrates the visualization of x-vectors using t-SNE \cite{tsne}. 
It can be observed that the speaker embeddings extracted using the x-vector model could discriminate different speakers.
For SID network, we choose two LSTM layers with 150 units each followed by a time distributed dense layer with 100 hidden units and softmax layer at the output. We choose categorical cross-entropy as the loss function for training the SID network. All the experiments in this work are performed using Keras \cite{Keras} with Tensorflow \cite{TF} as back-end.
To evaluate the performance of AAI-models we chose Root Mean Squared Error (RMSE) and Correlation Coefficient (CC) \cite{ghosh2010generalized, illa2018low}  as evaluation metrics computed for each articulator separately.
  
 \begin{table}[]
\centering
 	\vspace{-.2cm}
	\renewcommand\tabcolsep{2pt}
	
	\caption{RMSE and CC averaged across all the articulators and speakers (M1--M10 and F1--F10) with different AAI models in the closed-set (seen speakers) evaluation}
	\label{SeenAvgValsXv}

	\begin{tabular}{|c|c|c|c|c|c|}
		\hline
		& SD AAI                                                      & GM AAI                                                     & GM-FSD AAI                                                 & SC AAI                                            & xSC AAI                                           \\ \hline
		\begin{tabular}[c]{@{}c@{}}CC \\ (SD)\end{tabular}  & \begin{tabular}[c]{@{}c@{}}0.8361\\ (0.020)\end{tabular} & \begin{tabular}[c]{@{}c@{}}0.8608\\ (0.018)\end{tabular} & \begin{tabular}[c]{@{}c@{}}0.8699\\ (0.021)\end{tabular} & \begin{tabular}[c]{@{}c@{}}0.8721\\ (0.019)\end{tabular} & \begin{tabular}[c]{@{}c@{}}\textbf{0.8736}\\ (0.019)\end{tabular} \\ \hline
		\begin{tabular}[c]{@{}c@{}}RMSE\\ (SD)\end{tabular} & \begin{tabular}[c]{@{}c@{}}1.166\\ (0.076)\end{tabular}  & \begin{tabular}[c]{@{}c@{}}1.085\\ (0.072)\end{tabular}  & \begin{tabular}[c]{@{}c@{}}1.057\\ (0.083)\end{tabular}  & \begin{tabular}[c]{@{}c@{}}1.049\\ (0.077)\end{tabular}  & \begin{tabular}[c]{@{}c@{}}\textbf{1.044}\\ (0.076)\end{tabular}  \\ \hline
	\end{tabular}
	 	\vspace{-.4cm}
\end{table}
  
\section{Results and discussion}
In this section, we present the results of the experiments which compare the performance of xSC AAI with baseline AAI models in both seen and unseen speaker evaluations.


\textit{Seen speaker evaluations:}
Table \ref{SeenAvgValsXv} reports CC and RMSE using xSC AAI model and baseline models which are averaged across all the articulators and speakers.
From Table \ref{SeenAvgValsXv}, it is observed that xSC AAI performs better than SC AAI  which are followed by GM-FSD, GM, SD AAI models in terms of average CC and RMSE.
The performance of xSC AAI is on par with SC AAI and found to be consistent with the results reported in \cite{SCOHv}, where it has been shown that the SC AAI using one-hot encoded speaker conditioning perform better than all the baseline schemes in the closed-set condition.

\begin{table*}[htb]
	\centering
	
	\caption{CC averaged across all the unseen speakers (M11--M15 and F11--F15) for each articulator separately} 
	\label{UnseenArti}
	
	\begin{tabular}{|c|c|c|c|c|c|c|c|c|c|c|c|c|}
		\hline
		& $UL_x$ & $UL_y$ & $LL_x$ & $LL_y$ & $Jaw_x$ & $Jaw_y$ & $TT_x$ & $TT_y$ & $TB_x$ & $TB_y$ & $TD_x$ & $TD_y$ \\ \hline
		GM AAI   & 0.594  & 0.557  & 0.664  & 0.806  & 0.746   & 0.764   & 0.831  & 0.874  & 0.851  & 0.825  & 0.848  & 0.820  \\ \hline
		xSC AAI  & 0.603  & 0.602  & 0.691  & 0.806  & 0.740   & 0.771   & 0.838  & 0.876  & 0.856  & 0.836  & 0.854  & 0.828  \\ \hline
		SD AAI & 0.744  & 0.701  & 0.807  & 0.840  & 0.846   & 0.836   & 0.876  & 0.889  & 0.885  & 0.889  & 0.884  & 0.884  \\ \hline
	\end{tabular}
\end{table*}

\begin{table}[]
	\centering
	\caption{RMSE and CC averaged across all the articulators and speakers  (M11--M15 and F11--F15) with different AAI models}
	\label{UnseenXv}
	
	\begin{tabular}{|c|c|c|c|c|}
		\hline
		\multirow{2}{*}{} & \multicolumn{3}{c|}{Unseen}                                                                                                                                                    & Seen                                                      \\ \cline{2-5} 
		& GM AAI                                                      & uSC AAI                                                   & xSC AAI                                                    & SD AAI                                                   \\ \hline
		CC                & \begin{tabular}[c]{@{}c@{}}0.7655\\ (0.034)\end{tabular} & \begin{tabular}[c]{@{}c@{}}0.7534\\ (0.032)\end{tabular} & \begin{tabular}[c]{@{}c@{}}\textbf{0.7754}\\ (0.031)\end{tabular} & \begin{tabular}[c]{@{}c@{}}0.846\\ (0.027)\end{tabular}   \\ \hline
		RMSE              & \begin{tabular}[c]{@{}c@{}}1.4075\\ (0.108)\end{tabular} & \begin{tabular}[c]{@{}c@{}}1.4562\\ (0.100)\end{tabular} & \begin{tabular}[c]{@{}c@{}}\textbf{1.3933}\\ (0.100)\end{tabular} & \begin{tabular}[c]{@{}c@{}}1.1506\\ (0.1026)\end{tabular} \\ \hline
	\end{tabular}
\end{table}

\begin{figure}[htb]
	\centering
	\vspace{-.28cm}
	\centerline{\includegraphics[trim = 2mm 55mm 10mm 0mm, clip,width=9cm]{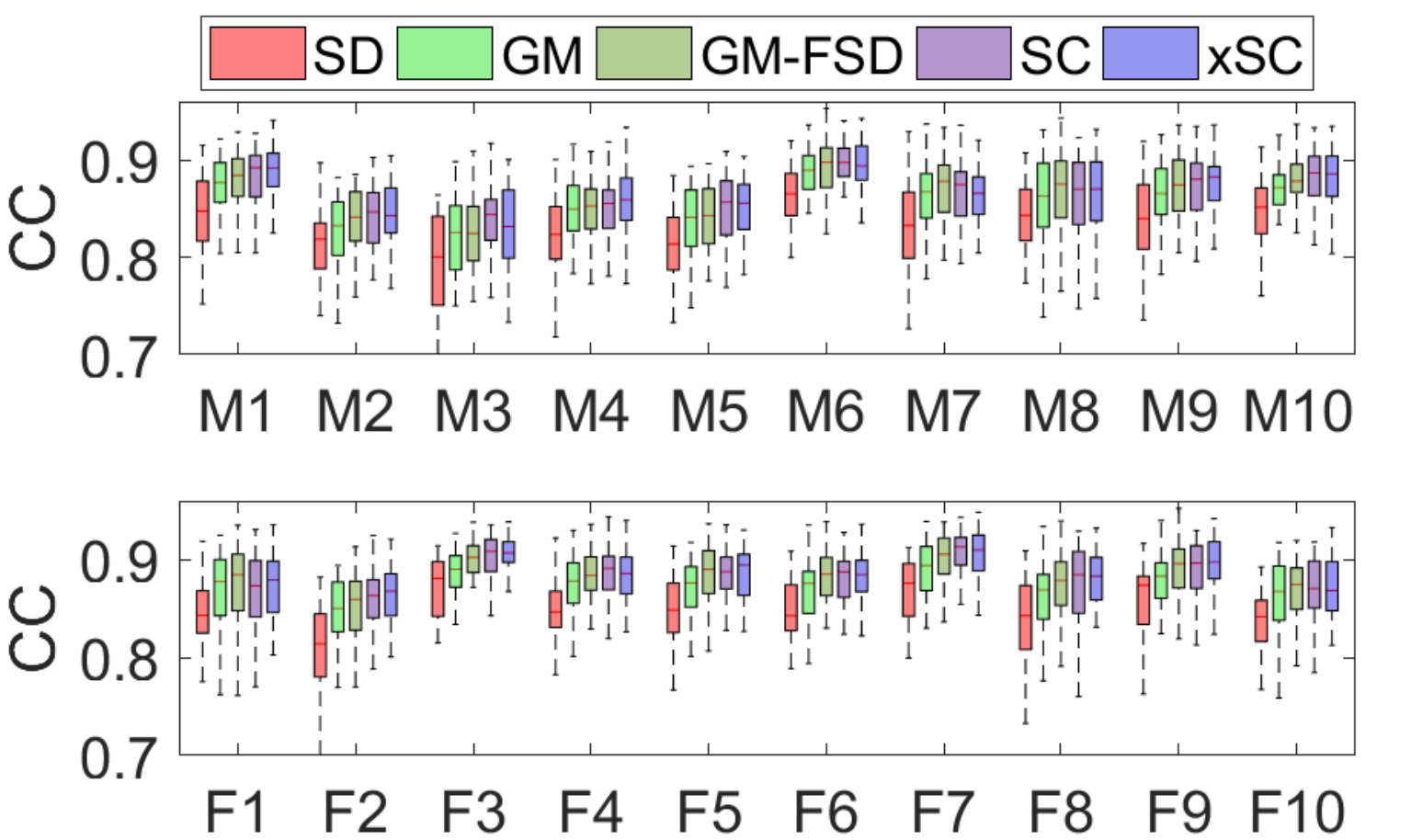}} 
	\vspace{-.35cm}
	\caption{CC averaged across all articulators for each of 20 speakers (male and female speakers in top and bottom rows respectively) in the closed-set (seen speakers) evaluation}
	\label{Seenspkplot}
	\vspace{-.4cm}
\end{figure}

 Speaker specific performance in terms of average CC with different AAI models are  shown in Fig. \ref{Seenspkplot} using boxplots, where in each box bottom edge represents the first quartile, and the line in the middle represents the median  and the upper edge of the box represents the third quartile.
 To examine the statistical significance in the performance difference between the xSC AAI model and the baseline models, we perform t-test on the CC values obtained from test sentences in a speaker specific manner.
 We observe that across all the speakers, the xSC AAI model performs significantly ($\rho < 0.05$) better than the SD AAI model.
 Similarly, while comparing xSC AAI with GM AAI, we observed significant ($\rho < 0.05$) improvement in performance across all the speakers except for M3 and M7.
 While comparing xSC AAI with GM-FSD, significant ($\rho < 0.05$) improvement is only observed in four speakers, namely, M5, F2, F3, and F8.  For the rest of the speakers, the performance of  xSC AAI is on par with GM-FSD. In spite of similar performance, the advantage of xSC AAI over GM-FSD would be that one single model of xSC AAI will capture multiple speakers' AAI mappings, unlike 20 separate AAI models in the case of GM-FSD.
 Similarly, comparing xSC AAI and SC AAI reveals that there is no statistical  ($\rho < 0.05$)  difference in performance except for M4, M9, and F2 speakers.

 \textit{Unseen speaker evaluations:}
 For unseen case evaluation, Table \ref{s2:dataset} presents the results for 10 speakers in terms of average CC and RMSE with different models.
 For unseen condition, we present the results of GM, xSC AAI, and uSC AAI models.
 We also report the performance on these 10 speakers in seen condition with the SD AAI model for reference comparison. 
 While comparing uSC AAI with GM AAI, we observe that there is a drop in performance with uSC AAI.
  Interestingly, we observe that using xSC AAI with x-vector performs better than the GM AAI.
  While comparing the performance of GM and xSC AAI models in the unseen case with respective SD AAI models, we observe there is a relative drop in CC of 10.94$\%$ and 9.51$\%$ for GM AAI and xSC AAI, respectively.
  This indicates that the drop in CC using xSC AAI model is $\sim$1.4$\%$ less than the GM AAI model.
  \begin{figure}[htb]
  	\centering
  	\centerline{\includegraphics[trim = 2mm 120mm 10mm 0mm, clip,width=9.5cm]{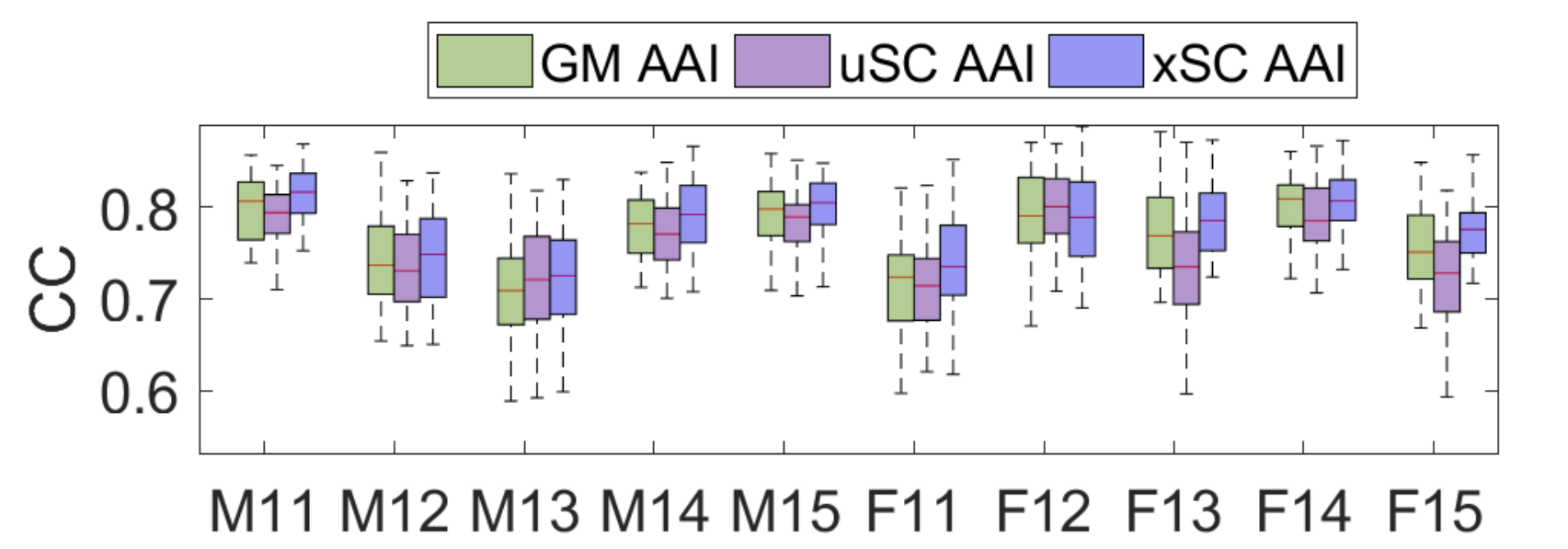}} 
  	\vspace{-.35cm}
  	\caption{CC averaged across all articulators for each of 10 speakers in unseen speaker evaluation}
  	\label{unseenCC}
  	\vspace{-.4cm}
  \end{figure}
Fig. \ref{unseenCC}, reports speaker specific analysis of the AAI performance in unseen case evaluation in terms of average CC across all the articulators.
We observe that the GM AAI performs better than the uSC AAI model.
The representation obtained from SID network in unseen speaker evaluation, utilized for uSC AAI model, could not generalize well. This might be due to few speakers (20) used for training SID network.
While comparing xSC AAI with GM AAI, we observe an improvement in performance in the majority of the speakers.
Similar to seen case condition, we perform analysis on results using a t-test to asses the statistical significance in the improvement of xSC AAI over the GM AAI model. We observe that there is a significant improvement in the performance using xSC AAI compared to GM AAI for majority of speakers, except for M12, F12, and F14.

\textit{Articulatory specific analysis:}
We also perform analysis to compare the performance of xSC AAI in an articulatory specific manner.
Table \ref{UnseenArti}, reports the performance of AAI models in terms of CC averaged across 10 unseen speakers. 
We also report SD AAI performance results of 10 speakers as a reference during the comparison.
We can observe that in subject independent AAI models (GM and xSC AAI) the drop in performance in CC compared to SD AAI are relatively less for tongue articulators followed by jaw and lips.
In unseen case evaluations, we observe that, in majority of the articulators, xSC AAI performs better than GM. We further perform a t-test for statistical significance. It is observed that, for all the articulators except $UL_x$, $LL_y$, and $TT_y$, there is a significant  ($\rho < 0.05$) difference in performance between GM and xSC AAI.

The improvements with xSC AAI could be due to the x-vectors which can encode speaker characteristics using TDNN trained with a large number of speakers in the VoxCelb dataset.

\section{Conclusions}

In this work, we experimentally study the benefit of x-vector conditioning in acoustic-to-articulatory inversion with multiple speakers' acoustic-articulatory data for training.
Experimental results revealed that there is no performance drop with x-vectors compared to one-hot representation in closed-set speaker evaluation.
In unseen speaker evaluation, xSC AAI using x-vectors generalizes well to the unseen speakers and performs better than the baseline uSC AAI and GM AAI models.  
A limitation of this work is the number of speakers available for xSC AAI training, as the acoustic-articulatory data from a large number of speakers is not available. 
In future, we will investigate the xSC AAI performance with respect to the number of speakers, and also with cross corpus evaluations. It is also interesting to study the performance of xSC AAI by including the training speakers from different native  languages and patients with speech disorders. 


\bibliographystyle{IEEEtran}

\bibliography{sampbib}


\end{document}